\documentclass[a4paper,11pt]{article}
\pdfoutput=1 

\usepackage{jheppub} 

\usepackage[T1]{fontenc} 

\newcommand{\be}{\begin{equation}}
\newcommand{\ee}{\end{equation}}
\newcommand{\bea}{\begin{eqnarray}}
\newcommand{\eea}{\end{eqnarray}}

\newcommand{\gapp}{\mathrel{\raise.3ex\hbox{$>$}\mkern-14mu \lower0.6ex\hbox{$\sim$}}}
\newcommand{\lapp}{\mathrel{\raise.3ex\hbox{$<$}\mkern-14mu \lower0.6ex\hbox{$\sim$}}}
\def\bbox{{\,\lower0.9pt\vbox{\hrule \hbox{\vrule height 0.2 cm
\hskip 0.2 cm \vrule  height 0.2 cm}\hrule}\,}}

\title{\boldmath Inconsistencies in Verlinde's emergent gravity}

\author[a]{De-Chang Dai}
\author[b]{Dejan Stojkovic }

\affiliation[a]{Institute of Natural Sciences, \\ Shanghai Key Lab for Particle Physics and Cosmology,\\ and Center for Astrophysics and Astronomy,\\ Department of Physics and Astronomy, \\ Shanghai Jiao Tong University, \\ Shanghai 200240, China}
\affiliation[b]{HEPCOS, Department of Physics, \\SUNY at Buffalo, \\ Buffalo, NY 14260-1500, USA}

\emailAdd{diedachung@gmail.com}
\emailAdd{ds77@buffalo.edu}

\abstract{We point out that recent Verlinde's proposal of emergent gravity suffers from some internal inconsistencies.
The main idea in this proposal is to preserve general relativity at short scales where numerous tests verified its validity, but modify it on large scales where we meet puzzles raised by observations (in particular dark matter), by using some entropic concepts. We first point out that gravity as a conservative force is very difficult (if possible at all) to portray as an entropic force. We then show that the derivation of the MOND relation using the elastic strain idea is not self-consistent. When properly done, Verlinde's elaborate procedure recovers the standard Newtonian gravity instead of MOND.}

\begin{document}
\maketitle
\flushbottom

\section{Introduction}

The idea of emergent gravity, put forward in \cite{Verlinde:2010hp,Verlinde:2016toy}, is an interesting attempt to attack the longstanding problems in gravity and cosmology from a new perspective. As such, it is very valuable since it is very unlikely that the problems we are facing will be resolved by some straightforward extension of the existing models. Then in \cite{Hossenfelder:2017eoh}, an explicit Lagrangian capturing some features of this proposal was proposed (see also \cite{Dai:2017guq} for important corrections). The purpose of this note is, however, to point out that, as it stands now, the proposal detailed in \cite{Verlinde:2010hp,Verlinde:2016toy} does not appear to be self-consistent.

\section{Gravity cannot be an entropic force}
\label{sec1}

We start with the first and most important premise of the entropic gravity proposal.  The main claim that relates entropy and gravitational force can be found around equation (3.7) in the first Verlinde's paper  \cite{Verlinde:2010hp}.

{\it
``The basic idea is to use the analogy with osmosis across a semi-permeable
membrane. When a particle has an entropic reason to be on one side of the membrane
and the membrane carries a temperature, it will experience an effective force equal to
\begin{equation}\label{ve}
F \triangle x = T \triangle S
\end{equation}
This is the entropic force." [$F$ is gravitational force, $\triangle x$ is the displacement, $T$ is the temperature of the system, and  $\triangle S$ is the change in entropy.]
}

This implies that the particle moves because the entropy of the system increases when it moves. However, this very first premise cannot be true. Namely, Newtonian gravitational force is conservative. An essential feature of conservative forces is that their action is always reversible. A system in a free fall will never increase its entropy, because this is a reversible process. We need some kind of dissipation (e.g. collisions) to increase entropy and make it irreversible. In general relativity this is realized by emission of gravitons which increase entropy. However, even in general relativity, it is possible to construct a freely falling (collapsing) system which does not radiate gravitons or any other radiation (for example a spherically symmetric case). Therefore gravity cannot be interpreted as an entropic force.

In fact, it is easy to see what is technically wrong with equation (3.7) in \cite{Verlinde:2010hp}.
The right hand side of that equation is work done by the gravity. The full expression should read
\begin{equation}\label{vec}
F \triangle x = T \triangle S +\triangle E_k
\end{equation}
where $\triangle E_k$ is the change in the kinetic energy of this system. Thus, gravitational force can perform work which results in the change of kinetic energy, while the entropy of the system remains constant all the time. In \cite{Verlinde:2010hp}, the author omitted $\triangle E_k$ and concluded that gravity was an entropic force.

Thus, gravity cannot be an entropic force, at least as long as entropy is the usual thermodynamical entropy as used in Eqs.~(\ref{ve}) and (\ref{vec}).

\section{Inconsistency in derivation of the MOND relation}
\label{sec2}

Recently, the original proposal was extended and refined in a new paper \cite{Verlinde:2016toy}. The main goal now is to solve the dark matter problem using the entropic gravity idea.
In section 7.1 of \cite{Verlinde:2016toy}, the author tries to prove that the surface mass density, $\Sigma_D$, for the apparent dark matter in terms of the Newtonian potential for the baryonic matter, $\Phi_B$, is (equation 7.37 in \cite{Verlinde:2016toy})
\begin{equation}
\label{MOND1}
\Big(\frac{8\pi G}{a_0}\Sigma_D\Big)^2 = \frac{d-2}{d-1}\triangledown_i \Big(\frac{\Phi_B}{a_0} n_i\Big)
\end{equation}
where  $d$ is the  number of dimensions in space, while $n_i$ is a normalized eigenvector satisfying $|n|^2 = 1$. The parameter $a_0$ is an acceleration scale determined by the Hubble constant, $H_0$, and the speed of light, $c$, as $a_0 = c H_0$.
If we take a point particle of mass $M$ as an example, the corresponding Newtonian potential is $\Phi_B =-\frac{GM}{r}$. From equation (\ref{MOND1}), one finds the surface mass density for dark matter as
\begin{equation}
\label{sur1}
\Sigma_D =\frac{\sqrt\frac{2Ma_0}{96\pi^2 G}}{r} .
\end{equation}
Since the surface mass density drops as $1/r$, the total mass grows with distance as $r$, so this behavior reproduces the MOdified Newtonian Dynamics (MOND), and can explain the flat galactic rotational curves at large distances.

However, to put the discussion in terms of the spacetime metric, the author introduces the displacement field, $u_i$, which is an analog of the gravitational potential, and the corresponding elastic strain tensor, $\epsilon_{ij}$, which is an analog of the gravitational acceleration. The displacement field is defined in equation 6.4 in \cite{Verlinde:2016toy} as
\begin{equation}
\label{displacement1}
u_i =\frac{\Phi_B}{a_0}n_i ,
\end{equation}
while the linear strain tensor is defined in equation 6.1 in \cite{Verlinde:2016toy} as
\begin{equation}
\epsilon_{ij}=\frac{1}{2}(\triangledown_i u_j +\triangledown_ju_i) .
\end{equation}
In the point particle case, the strain must be proportional to $1/r^2$, because it is just a derivative of $u_i$.

In equation (7.28) in \cite{Verlinde:2016toy}, the author writes down the relation between the apparent dark matter surface density and the principal strain $\epsilon(r)$
\begin{equation}
\label{sur12}
\Sigma_D =\frac{a_0}{8\pi G}\epsilon(r)
\end{equation}
where $\epsilon(r)$ is defined as
\begin{equation}
\label{ps}
\epsilon(r)n_i = \left( \epsilon_{ij} -\frac{1}{d-1}\epsilon_{kk} \delta_{ij} \right) n_j .
\end{equation}
If the principal strain $\epsilon(r)$ has the same general behavior as the strain tensor $\epsilon_{ij}$, i.e. falls off as $1/r^2$, then
 \begin{equation}
\label{sur2}
\Sigma_D =\frac{a_0}{8\pi G}\epsilon(r) \sim r^{-2} .
\end{equation}
This is very different from the desired form in equation (\ref{sur1}). Since equations (\ref{sur1}) and (\ref{sur2}) cannot be both right at the same time, the author goes through an elaborate construct to justify his choices. Basically, he is trying to force $\epsilon(r)$ to drop as $1/r$ rather than $1/r^2$. We will now go through the main steps, pointing out a major problem.

The main observation that the author utilizes in \cite{Verlinde:2016toy} is that the presence of ordinary matter in some subregion $\mathcal{B}$ of the de Sitter space removes the amount of entropy $S_M(\mathcal{B})$ out of the total de Sitter entropy. The removed entropy  is proportional to the displacement $u_i$ as
\begin{equation}
 S_M(\mathcal{B})=\frac{1}{V_0^*}  \int_{\partial \mathcal{B}}u_i dA_i
\end{equation}
where the integral goes over the area of the subregion $\mathcal{B}$. Here $V_0^*$ is a constant normalization term. The author expects that a point mass removes entropy within radius $r$ as
\begin{equation}
S_M(r)=-\frac{2\pi M r}{\hbar}
\end{equation}
This is consistent with $u_i \sim 1/r$, and justifies equation (\ref{displacement1}).  Next, the author proposes that the removed entropy is not perfectly spherically distributed. Ordinary matter is located in many smaller ``inclusion regions" labeled $\mathcal{V}_M (L)$. The displacement field now satisfies
\begin{equation}
\label{dark}
\triangledown_i u_i =
  \begin{cases}
    -V_0^*/V_0       & \quad \text{  inside  } \mathcal{B}\cap \mathcal{V}_M(L)\\
    0  & \quad \text{  outside  } \mathcal{B}\cap \mathcal{V}_M(L)\\
  \end{cases}
\end{equation}

\begin{figure}
   \centering
\includegraphics[width=9cm]{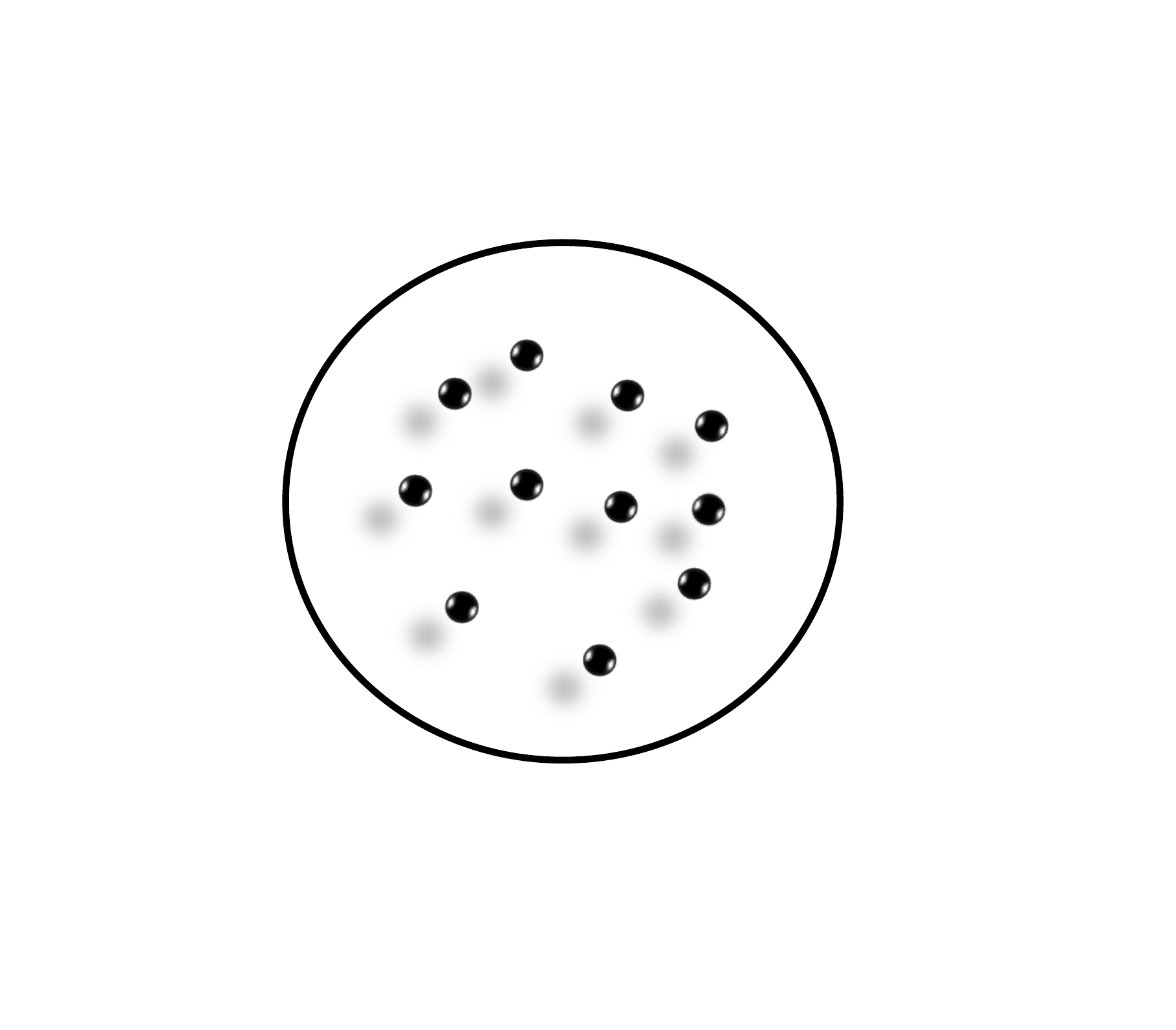}
\caption{The black solid circles are the inclusion regions, $\mathcal{V}_M (L)$, where matter is present. Within these regions entropy is removed, and the displacement satisfies $\triangledown_i u_i=-V_0^*/V_0$.  Outside of these regions entropy is not removed, and the displacement satisfies $\triangledown_i u_i=0$.
}
\label{entropy}
\end{figure}

Since these inclusion regions are randomly scattered in space (as in figure \ref{entropy}), the displacement $u_i$ is not exactly the same as in equation (\ref{displacement1}). It must be corrected to
\begin{equation}
\label{displacement2}
u_i =\frac{\Phi_B}{a_0}n_i +\vec{\delta}(x,y,z)
\end{equation}
where $\vec{\delta}(x,y,z)$ is the fluctuation caused by the non-uniform  distribution of removed entropy regions. If the fluctuations are random, then in large areas they cancel out on average, or at least they are much smaller than the first term, $\frac{\Phi_B}{a_0}n_i$, i.e.
\begin{equation}
|\int \vec{\delta}(x,y,z) \cdot d\vec{A}|\ll |\int \frac{\Phi_B}{a_0}n_i dA_i| .
\end{equation}
In this context, averaging means integration.
The random distribution of inclusion regions also modifies the strain as
\begin{equation}
\label{strain-o}
\epsilon (r)= \frac{H}{r^2} +f(x,y,z)
\end{equation}
where $H$ is some constant which is not very important, while $f(x,y,z)$ is the fluctuation in the strain induced by the fluctuation in the displacement $\vec{\delta}(x,y,z)$. We can now compute the volume integral of $\epsilon (r) ^2$ as
\begin{equation}
\label{strain-in}
\int \epsilon (r)^2 dV = \int \frac{H^2}{r^4} dV +2\int \frac{H}{r^2}f(x,y,z) dV + \int f(x,y,z)^2 dV .
\end{equation}
The term $\int \frac{H^2}{r^4} dV$ is the standard average behavior. The term linear in $f(x,y,z)$, i.e. $\int \frac{H}{r^2}f(x,y,z) dV$ will be canceled out on average. However, the term quadratic in fluctuations,  $\int f(x,y,z)^2 dV$,  survives and corresponds to an additional contribution to the strain from the random distribution of inclusion regions. The author expects that on average
\begin{equation}
\label{flow}
\int \epsilon (r) ^2 dV = \frac{d-2}{d-1}\int_{\partial \mathcal{B}} u_i dA_i =\frac{d-2}{d-1}\int_{\partial\mathcal{B}}\frac{\Phi_B}{a_0}n_i d A_i .
\end{equation}
In the last step, the fluctuation $\vec{\delta}$ in equation (\ref{displacement2}) is removed after averaging. Thus, from equations (\ref{strain-in}) and (\ref{flow}) one gets
\begin{equation}
\int \frac{H^2}{r^4} dV+\int f(x,y,z)^2 dV \approx \frac{d-2}{d-1}\int_{\partial\mathcal{B}}\frac{\Phi_B}{a_0}n_i d A_i
\end{equation}
As already mentioned, the term $2\int \frac{H}{r^2}f(x,y,z) dV$ is removed because it is linear in fluctuations. $\int \frac{H^2}{r^4} dV$ is the regular term, which decays like $r^{-1}$. To obtain the correct   right hand side which is proportional to $r$, the extra contribution term $\int f(x,y,z)^2 dV $ must grow like $r$. This implies that the fluctuation $f(x,y,z)$ falls off as $1/r$.  The singularity at $r=0$ is neglected, because point particle description will fail at some small finite radius. Using the relation in equation (\ref{sur12}), the author also rewrites equation (\ref{flow}) in terms of the surface density (it is equation 7.36 in \cite{Verlinde:2016toy}) as
\begin{equation}
\int _B \Big(\frac{8\pi G}{a_0 }\Sigma_D\Big)^2 dV = \frac{d-2}{d-1}\int_{\partial\mathcal{B}}\frac{\Phi_B}{a_0}n_i d A_i .
\end{equation}
One then applies Stokes theorem to obtain the crucial equation (\ref{MOND1}). This is how Verlinde derives the MOND  relation in section 7.1 of \cite{Verlinde:2016toy}.

What is wrong with this procedure? The main technical point is that averaging was applied  to remove contribution from the fluctuation $\vec{\delta}$, or equivalently $2\int \frac{H}{r^2}f(x,y,z) dV$ in equation (\ref{strain-in}). So again, after integration of equation (\ref{strain-o}) one gets
\begin{equation}
\label{strain-in-1}
\int \epsilon (r) ^2 dV \approx \int \frac{H^2}{r^4} dV+\int f(x,y,z)^2 dV .
\end{equation}
This is all that one can conclude at this point.
By equating the terms under the integral, one can naively find that
\begin{equation}
\epsilon(r) \approx \sqrt{\frac{H^2}{r^4}+f(x,y,z)^2} .
\end{equation}
Since $f(x,y,z)$ falls off as $1/r$, at large distances the strain and thus the surface density $\Sigma_D$ (because of equation (\ref{sur12})) falls off as $1/r$, exactly as needed for MOND. However, this is incorrect. The cross term in equation (\ref{strain-in}) is gone only after the integration because of the averaging, so one cannot go back and extract $\epsilon(r)$ this way. For the same reason equation (\ref{flow}) cannot be applied to recover $\epsilon(r)$, because $\vec{\delta}$ is removed by averaging.

To avoid this mistake, one has to go back to equation (\ref{strain-o}). On average, the surface density is

\begin{eqnarray}
&& \Sigma_D =\frac{1}{A} \int \frac{a_0 }{8\pi G} \epsilon (r) dA =  \\
&& \frac{1}{A} \frac{a_0 }{8\pi G} \int \left[ \frac{H}{r^2} +f(x,y,z)\right] dA\approx   \frac{1}{A} \frac{a_0 }{8\pi G} \int  \frac{H}{r^2}  dA \nonumber,
\end{eqnarray}
where $A$ is the area over which one integrates. The linear term $\int f(x,y,z) dA$ is again suppressed. The final result is that the surface density falls off as $1/r^2$, not as desired $1/r$. This is the same behavior as in Newtonian gravity, not MOND. Therefore, the crucial MOND relation written in equation (\ref{MOND1}) cannot be self-consistently derived in this way.

\section{Conclusions}

In this paper we pointed out two major problems that plague Verlinde's proposal of emergent gravity. While the proposal contains some attractive features, a self-consistent formulation (if possible at all) requires addressing the problems we outlined here. The first problem is that the equation (\ref{ve}) on which the entropic reasoning is based in incomplete, and instead  equation (\ref{vec}) should be used. Then it will become clear that gravity as a conservative force cannot have an entropic origin. The second problem appears when an attempt was made to derive the MOND relation in equation (\ref{MOND1}).  The averaging procedure was not applied appropriately, and instead of the regular mean the root mean square was used. We showed that when the averaging is properly done, the contribution from the strain behaves like ordinary Newton's gravity instead of MOND.

Entropic reasoning from the section \ref{sec1} was also criticized in \cite{Visser:2011jp}, where Eq.~(\ref{ve}) was generalized to multiple heat baths with multiple
temperatures and multiple entropies. However, a more complete Eq.~(\ref{vec}) was not discussed in \cite{Visser:2011jp}.

It is also instructive to note that our discussion from the section \ref{sec1} does not apply (at least not in a straightforward way) to different approaches that can be found in the literature. For example, it was argued in \cite{Jacobson:1995ab,Padmanabhan:2002ma} that a thermodynamic interpretation of the relativistic Einstein equations might be possible (as opposed to the Newtonian force like in Verlinde's proposal). However, neither of these proposals is explicitly using the form of Eq.~(\ref{ve}). In particular, in \cite{Jacobson:1995ab}, to argue that the Einstein equations are an analog of a thermodynamical equation of state, the relation $\delta Q = T dS$ was used, which is technically correct, while Eq.~(\ref{ve}) is incomplete. In \cite{Padmanabhan:2002ma}, the author gives an interpretation that the Einstein's gravitational action represents the free energy of the spacetime geometry. Since this interpretation does not involve any incomplete thermodynamical relations, our criticism does not apply to it.

Our discussion from the section \ref{sec2} does not apply to earlier approaches in \cite{Jacobson:1995ab,Padmanabhan:2002ma} since it crucially depends on the elaborate procedure in \cite{Verlinde:2016toy}. For some recent approaches to emergent spacetime see also \cite{Afshordi:2014cia,Edmonds:2017zhg,Cadoni:2017evg}.

\begin{acknowledgments}
D.C Dai was supported by the National Science Foundation of China (Grant No. 11433001 and 11447601), National Basic Research Program of China (973 Program 2015CB857001), the key laboratory grant from the Office of Science and Technology in Shanghai Municipal Government (No. 11DZ2260700) and  the Program of Shanghai Academic/Technology Research Leader under Grant No. 16XD1401600. D.S. was partially supported by the US National Science Foundation, under Grant No. PHY-1417317.
\end{acknowledgments}

\end{document}